\documentclass[draft]{agujournal2019}
\usepackage{url} 
\usepackage[inline]{trackchanges} 
\usepackage{soul}
\usepackage{amsmath}

\usepackage{appendix}

\newcommand{\tiu}{J~m$^{-2}$~K$^{-1}$~s$^{-1/2}$}

\newcommand{\Lsp}{L$_\text{sp}$}
\newcommand{\mum}{$\mu$m}

\draftfalse

\journalname{JGR: Planets}

\begin{document}

%
%


\title{On the Possibility of Melting Water Ice during the Recent Past of Mars: Implications for the Formation of Gullies}

%
%




\authors{L.Lange\affil{1},  F.Forget\affil{1}}

\affiliation{1}{Laboratoire de Météorologie Dynamique, Institut Pierre-Simon Laplace (LMD/IPSL), Sorbonne Université, Centre National de la Recherche Scientifique (CNRS), École Polytechnique, École Normale Supérieure (ENS), Paris, France}
\correspondingauthor{Lucas Lange}{lucas.lange@lmd.ipsl.fr}



\begin{keypoints}
\item Melting of opaque water snow or ice at the surface of Mars over the last 4 million years is prevented by sublimation cooling.
\item Melting of subsurface ice in equilibrium with the atmosphere is also unlikely because it is too deep to be warmed enough. 
\item Subsurface ice brought near the surface by regolith removal is unlikely to melt, as it happens only under unrealistic conditions.
\end{keypoints}

%
%

%
%
 \begin{abstract} 
The formation of gullies on Mars has often been attributed to the melting of (sub)surface water ice. However, melting-based hypotheses generally overlook key processes: (1) sublimation cooling by latent heat absorption, (2) the non-stability of ice where melting conditions can be reached, and (3) the particular microclimates of gullied slopes. Using state-of-the-art climate simulations, we reassess ice melting scenarios over the past four million years (obliquity $\le$35\textdegree), beyond the estimated period of gully formation. We find that the melting of opaque water snow or ice at the surface of Mars is unlikely anywhere due to sublimation cooling, while (quasi-) stable subsurface ice is typically too deep to reach melting temperatures. We propose an alternative mechanism in which seasonal CO$_2$ frost sublimation destabilizes the regolith and brings the underlying water ice close to the surface, allowing rapid heating. Even under these optimal conditions, melting requires unrealistic assumptions. Ice containing a small amount of dust could melt via a solid-state greenhouse effect, but both its possibility and frequency in Mars' recent past remain uncertain.
\end{abstract}

\section*{Plain Language Summary}
Mid-latitude slopes on Mars display young surface features known as gullies, which resemble small terrestrial alluvial fans. Owing to this similarity, they were considered evidence for recent liquid water activity, within the last few million years. Several formation models invoke the melting of surface or subsurface water ice. However, these models often neglect three key processes: (1) sublimation consumes latent heat, significantly cooling the ice; (2) water ice is unstable under conditions where melting could occur, since it sublimates first; and (3) gullies typically occur on pole-facing slopes, which are generally too cold to reach melting temperatures. Using new numerical models that account for these effects, we reinvestigate whether ice could have melted recently on Mars. Our results show that fresh snow or opaque ice at the surface cannot melt due to sublimation cooling, and subsurface ice lies too deep to be sufficiently warmed. The most favorable scenario involves ice approaching the surface as seasonal CO$_2$ frost sublimation erodes the overlying soil. Although this could cause transient heating, melting only occurs under unrealistic assumptions.   Slightly dusty ice exposed at the surface could melt at depth, yet the frequency and conditions of its occurrence in Mars' recent past are unknown.

\section{Introduction}
The detection of liquid water on a planet is of major importance, as its presence on Earth is systematically associated with the existence of life \cite{Rothschild2001}. Evidence for past liquid water on Mars comes from dendritic valley networks \cite<e.g.,>[]{Carr1981}, widespread hydrated minerals \cite{Poulet2005, Carter2023}, and delta-lake systems and fluvial sedimentary conglomerates \cite{Malin2003, Williams2013, Mangold2021}. Yet, most of these records date from the Noachian to early Amazonian eras ($>$~3 billion years), leaving debate about recent liquid water \cite{Fassett2008}. Although present-day martian surface pressure and temperature can locally exceed the triple point of water (611~Pa, 273.15~K) \cite{Haberle2001}, the atmospheric water vapor content remains extremely low \cite<$\sim$10~pr-$\mu$m, corresponding to a vapor partial pressure of $\sim$ 0.1~Pa, well below saturation,>{Smith2002}.As a consequence of this very low relative humidity (typically of order 0.01\%), surface liquid water would undergo rapid evaporation or boiling under typical conditions and cannot be persistently sustained in the absence of continuous replenishment, in contrast to terrestrial environments where evaporative losses are compensated by an active hydrological cycle.

Yet geological observations \cite<see review in>[]{Conway2021}  may suggest the presence of liquid water on the surface of Mars in a recent past. Notably, gullies, young \cite<younger than 4 million years>[]{deHaas2019} surface features found on slopes at mid and high latitudes \cite{Malin2000} composed of an alcove, a channel, and a depositional apron, are similar to small terrestrial alluvial fans. This analogy initially suggested that these surface features were formed by liquid water \cite{Malin2000}. Since then, many models have suggested that water ice melting at the surface \cite{Malin2000, Hecht2002, Christensen2003gullies, Kossacki2004, Williams2008, Dickson2023} or subsurface \cite{Costard2002, Kreslavsky2003, Morgan2010} could trigger the formation/activity of gullies.  As observations showed that activity occurs from some gullies during winter \cite{Malin2006, Dundas2022}, formation processes based on the sublimation of CO$_2$ ice have been shown to operate today rather than those based on water \cite<see>[]{Pilorget16, Dundas2019gullies}. However, the question remains whether these CO$_2$-based mechanisms are responsible for the formation of all gullies, given the low number of active gullies today. 
Recently, \citeA{Dickson2023} proposed that Martian gullies may have formed through the melting of water ice within the last million years, with present-day activity driven by the sublimation of CO$_2$ ice. Using climate simulations at 30\textdegree~and 35\textdegree~obliquity \cite<values corresponding to the highest obliquities reached during the last four million years>[]{Laskar2004}, they showed that gullies occur in regions where surface soil temperatures exceed 273~K and where the total atmospheric pressures surpass 612~Pa, owing to enhanced surface pressure from the sublimation of CO$_2$ ice deposits buried at the South Pole during high-obliquity periods \cite{Phillips2011}. Based on these thermal and pressure conditions, they concluded that gullies are associated with environments where melting could potentially occur. However, this scenario faces two main limitations:

\begin{enumerate}
\item The calculations by \citeA{Dickson2023} did not account for the cooling effect associated with sublimation. As shown by \citeA{Ingersoll1970}, even if there is no wind to transport water vapor away, the inevitable density contrast between water vapor and the main atmospheric constituent CO$_2$ is so strong that water vapor cannot accumulate above subliming ice, maintaining a sublimation rate that induces a significant latent heat cooling which prevents the ice temperature from reaching 273.15~K. This conclusion has been confirmed by \citeA{Schorghofer2020} and \citeA{Khuller2024}. This last study included the effects of the vapor/CO$_2$ density contrast, gustiness/large-scale wind, and the surface layer's specificity in their water ice sublimation modeling.

\item  A source of water, such as subsurface or surface ice, must be available for melting to occur. While \citeA{Dickson2023} emphasize that water frost and ground ice are generally observed in terrains where gullies form, we show in this paper that, under recent Martian conditions, water ice is almost inevitably lost wherever temperatures approach 273~K.

\end{enumerate}

Hence, the question of whether melting could have occurred on Mars in the recent past, and whether it could have contributed to the formation of gullies, remains open. This paper aims to investigate, using state-of-the-art Martian climate models, whether the melting of water ice at the surface/subsurface could occur at gully locations in the last 4 million years when the obliquity was $\le$~35\textdegree. The models used are presented in section \ref{sec:meltmethods}. The possibility of melting ice on the surface and in the subsurface is discussed respectively in section \ref{sec:surfacemelt} and \ref{sec:stableicetemp}. We discuss the possibility of melting subsurface ice after erosion of the near-surface in section \ref{ssec:meltunstable}.  Discussions are led in section \ref{sec:discussmelt} and conclusions are drawn in section \ref{sec:conclusionsmelt}.

\section{Methods}
\label{sec:meltmethods}
\subsection{Surface-Atmosphere model}

This study uses the 1D and 3D versions of the Mars Planetary Climate Model \cite<PCM,>{Forget1999}, along with the sub-grid scale slope microclimate parameterization outlined in \citeA{Lange2023Model}. For the 3D simulations, surface properties (albedo, emissivity, thermal inertia) are set to the observations from the Thermal Emission Spectrometer  \cite<TES,>[]{Putzig2007}. A constant dust opacity of $\tau$~=~0.2 is used (a relatively low value that favors higher daytime surface temperature). Paleoclimate simulations at high obliquity include the effect of radiative active clouds \cite{Madeleine2014}. For such simulations, we set the surface pressure to double its present-day value, assuming that CO$_2$ ice buried beneath the South Pole had entirely sublimed \cite{Phillips2011}. This favors the formation and stability of liquid water.  We test three configurations: present-day orbital configuration, 35\textdegree~obliquity with solar longitude of the perihelion \Lsp~=~90\textdegree, and 35\textdegree~obliquity with \Lsp~=~270\textdegree. The eccentricity is set to its present value, as all $\sim$35\textdegree~high-obliquity periods in the last 4 million years occurred with lower eccentricity.

To test the sensitivity of our results, we also used a 1D version of this model derived from the 3D version. The 1D model computes the same physical processes as in the 3D model, but the large-scale dynamics is not modeled. In this model, the surface properties are  "free parameters". 
 In our baseline study, we use an albedo of 0.13, a near-surface thermal inertia of 350~\tiu, and an emissivity of 1, consistent with observations at gully locations \cite{Harrison2015}. In addition, sensitivity tests were performed by varying the thermal inertia between 50 and 350~\tiu and the surface albedo up to 0.3, a range representative of martian bare-surface properties \cite{Putzig2007}. As discussed in section \ref{ssec:meltunstable}, lower thermal inertia reduces the amount of energy conducted from the surface to subsurface ice, thereby decreasing subsurface heating and the likelihood of melting. A similar effect is obtained for albedo, as higher values reduce the amount of solar energy absorbed at the surface (not shown). A large-scale wind of 10~m~s$^{-1}$ is prescribed above the boundary layer \cite<a typical value used in 1D studies, e.g.,>[]{Savijrvi2024}, and the wind profile is calculated by the model \cite{Colaitis2013}.

\begin{figure}[h!]
  \centering
  \includegraphics[width = \textwidth]{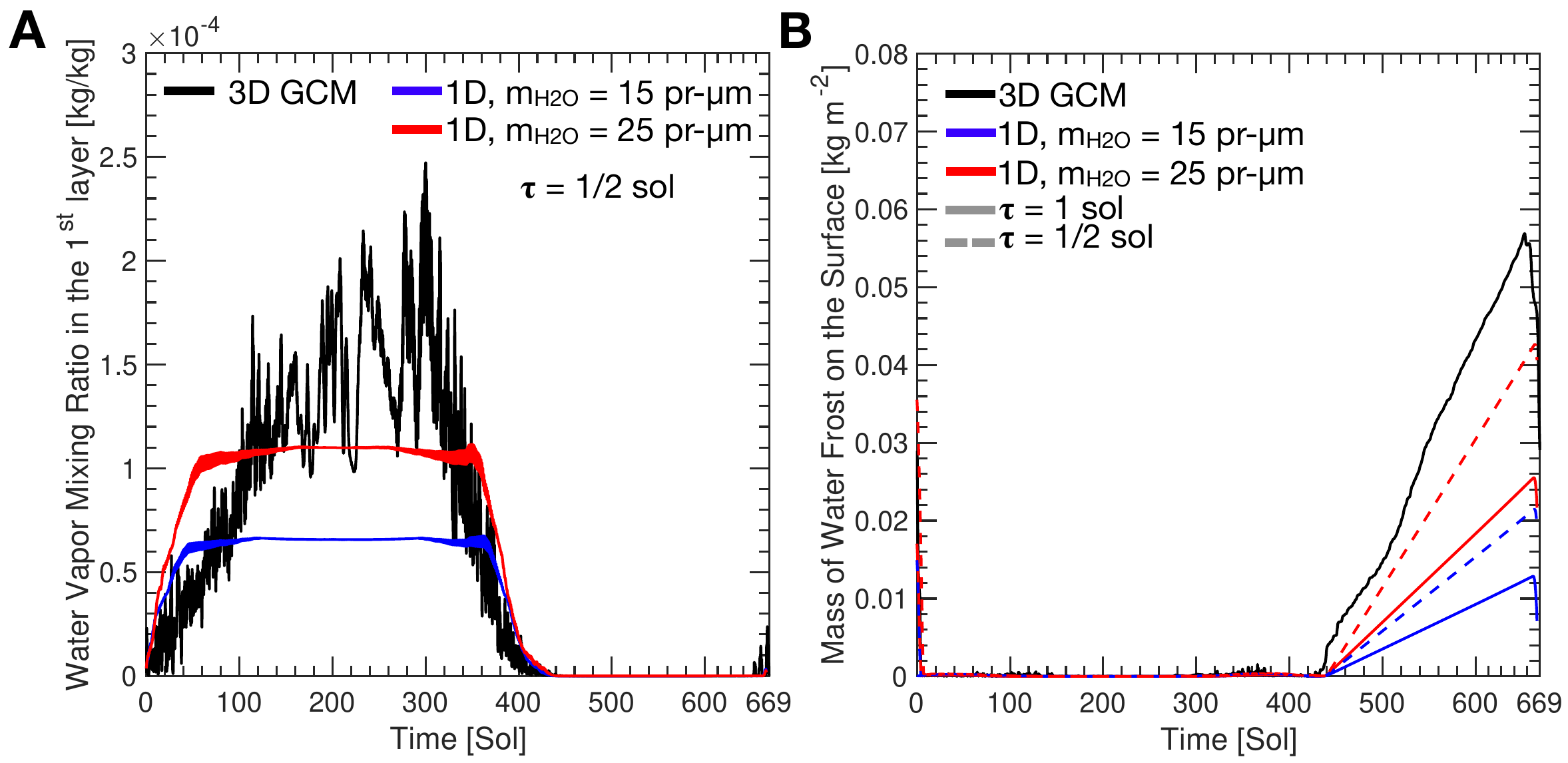}
  \caption{a) Mass mixing ratio of the water vapor in the first layer ($\sim$4 m) of the model (black curves) at a latitude of 56.25\textdegree N (longitudinally averaged), for $m_{\rm {H}_2\rm {O}}$~=~15~pr-$\mu$m (blue curve) and $m_{\rm {H}_2\rm {O}}$~=~25~pr-$\mu$m (red curve). In these simulations, $\tau$ is set to half of a Sol duration. b) Frost thickness predicted by the 3D and the 1D, for the same $m_{\rm {H}_2\rm {O}}$ as a), and for several relaxation times (one sol for the plain curve, half a sol for the dashed lines). }
  \label{fig:relaxation1D}
\end{figure}

As explained in the following sections, the sublimation of water ice, and hence latent heat cooling, is highly sensitive to the near-surface water vapor content. The latter depends strongly on large-scale meteorology and transport processes \cite{Montmessin2017}, which are resolved in the 3D model but not in the 1D model. We therefore represent the lateral transport of atmospheric water in the 1D model by applying a nudging toward a prescribed mean value, while the vertical transport is handled by the boundary-layer and diffusion schemes described in \citeA{Forget1999} and \citeA{Colaitis2013}. The mixing ratio of water vapor $q_{\rm{vap}}$~(kg~kg$^{-1}$) given as input to the model  at time $t + \Delta t$~(s) is:

\begin{equation}
    q_{\rm{vap}}(t + \Delta t) = \underbrace{\frac{m_{\rm {H}_2\rm {O}}g}{P_{\rm{surf}}}}_{\rm{average~mass~mixing~ratio~in~the~column}} + \left(q_{\rm{vap, model}}(t + \Delta t) - \frac{m_{\rm {H}_2\rm {O}}g}{P_{\rm{surf}}} \right)e^{-\frac{\Delta t}{\tau}}
\end{equation}

\noindent where $m_{\rm {H}_2\rm {O}}$~(kg~m$^{-2}$) is the nudging value of the H$_2$O mass in the column per unit area, $g$~=~3.72~m~s$^{-2}$ the gravity, $P_{\rm{surf}}$~(Pa) the surface pressure, $q_{\rm{vap, model}}(t + \Delta t)$ is the value of the water vapor mixing ratio predicted by the 1D model at time $t + \Delta t$ ($\Delta t$ being the timestep of the model), and $\tau$~(s) is the relaxation time. We also assume that clouds are advected away from the atmospheric column after each time step, so the mixing ratio of water ice in the air is always set to zero before entering a new time step.
To determine the nudging parameters ($m_{\rm {H}_2\rm {O}}$, $\tau$), we compared the water vapor in the first atmospheric layer of the model (~4 m) and the amount of water frost on the surface between the 1D model, for several nudging values, and the 3D model (with similar surface conditions - pressure, surface optical/thermophysical properties). An example is given in Figure \ref{fig:relaxation1D} for latitude 56.25\textdegree N where two values of $m_{\rm {H}_2\rm {O}}$ were tested: 15 pr-$\mu$m, the yearly average value measured by the Thermal Emission Spectrometer TES at this latitude \cite{Smith2002}, and 25 pr-$\mu$m, the maximum value measured by TES over the year at this latitude. Comparison with the 3D model shows that it is necessary to use the maximum value over the year measured by TES at a given latitude to represent the annual water cycle better.  By testing several values of the relaxation time $\tau$, we find that $\tau~=~1/2$~sol and $\tau~=~1/4$~sol (not shown in Figure \ref{fig:relaxation1D}) allow for a realistic representation of the seasonal cycle of water vapor in the atmosphere and frost at the surface, in terms of amplitude and temporal evolution.  A value of $\tau~=~1/2$~sol is therefore adopted in the simulations presented here.

By testing several values of the relaxation time $\tau$, it appears that choosing a relaxation time of 1/2 sol or 1/4 sol (not shown in Figure \ref{fig:relaxation1D}) is optimal. 

\subsection{Surface ice/frost sublimation model}
\subsubsection{General description:}
The sublimation of water frost/ice on the surface $F_{\text{ice}}$~(kg~m$^{-2}$~s$^{-1}$) is given by \cite{Lange2024}:

\begin{equation} 
     F_{\text{ice}} = C_q U \left( \rho_{\rm{vap,atm}} - \rho_{\rm{sv}}(T_{\text{ice}}) \right)  
\end{equation}

\noindent where $C_q$~(unitless) is a moisture transfer coefficient, $U$ (m~s$^{-1}$) is the wind velocity obtained by combining the large-scale (synoptic) wind near the surface with a wind gustiness induced by buoyancy \cite{Colaitis2013}, $\rho_{\rm{vap,atm}}$~(kg~m$^{-3}$) the water vapor density in the near-surface layer (altitude of $\sim$4~m), $\rho_{\rm{sv}}(T_{\text{ice}})$~(kg~m$^{-3}$)  the water vapor density at the icy surface derived from the ice temperature $T_{\text{ice}}$~(K)  \cite{Murphy2005}. C$_q$  depends on the stability of the atmosphere \cite{Colaitis2013, Khuller2024,Lange2024} and is given by:

\begin{equation}
    C_q = f_q(Ri) \left( \frac{\kappa^2}{\ln{\frac{z_1}{z_0}}  \ln{\frac{z_1}{z_{0q}}}}\right)
\end{equation}

\noindent where $f_q(Ri)$~(unitless) is a function of the Richardson number $Ri$~(unitless),  $\kappa$~(unitless) is the von Kármán constant set to 0.4; $z_0$~(m) is the aerodynamic roughness coefficient extracted from \citeA{Hebrard2012},  $z_{0q}$ (m) the moisture roughness length derived from the Reynolds and Prandtl numbers \cite<see Eq. 6 of>[]{Lange2024}. The stability function $f_q(Ri)$, given by \citeA{England1995}, amplifies the flux for an unstable atmosphere, while it reduces it for a stable atmosphere. The Richardson number $Ri$ is the ratio of the buoyancy term to the flow shear term, and depicts the stability of the atmosphere: 

\begin{equation}
    Ri = \frac{g z_1}{\theta_{\rm{v,s}}} \frac{(\theta_{\rm{v}}(z_1) - \theta_{\rm{v,s}})}{U^2}
    \label{eq:turb_rib}
\end{equation}

\noindent where $\theta_{\rm{v,s}}$~(K) is the virtual surface temperature, $\theta_{\rm{v}}(z_1)$~(K) is the virtual temperature at altitude $z_1$. This quantity \cite<see expression in Eq. 8 of>[]{Lange2024} represents the temperature that dry air with the same density and pressure as moist air would have. $Ri$~=~0 for a neutral atmosphere, $Ri~<~0$ for an unstable atmosphere, $Ri~>~0$ for a stable atmosphere. Hence, using virtual temperature, the stability of the atmosphere, and thus the sublimation of water ice, is determined by 1) the contrast of density between the near-surface air and atmosphere induced by the difference of temperature and 2) the contrast of density between the moist air at the surface where ice is subliming, and the atmosphere mainly composed of CO$_2$. While the latter effect is included in the Richardson number and the stability function in our model, we do not include it when computing the gustiness' wind. This is left as future work as it would require significant changes in the Planetary Boundary Layer within the Mars PCM.

\subsubsection{Model validation}

\begin{figure}[h!]
  \centering
  \includegraphics[width = \textwidth]{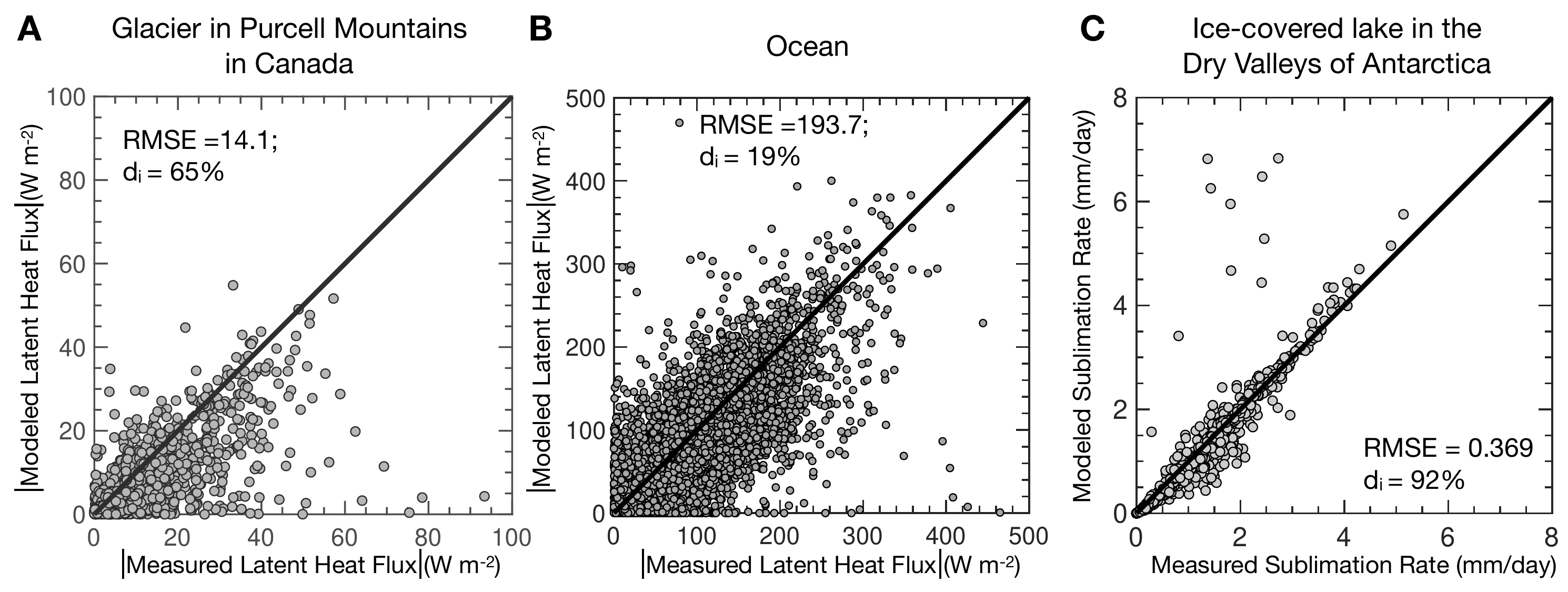}
  \caption{Comparison between modeled vs measured absolute latent heat flux (a, b) and sublimation rate (c) for a) ice glacier in Purcell Mountains \cite{Fitzpatrick2017}, b) oceanic surfaces \cite{Fairall1996, Fairall2003}, and c) an ice-covered lake in the Dry Valleys of Antarctica \cite{Clow1988}  The black line indicates perfect agreement between measured and modeled values. For each case, the root mean square error and dimensionless index of agreement ($d_i$) \cite<see section 5 of>{Khuller2024} are given. }
  \label{fig:validation}
\end{figure}

Our sublimation model is tested against the latent heat-flux measurements on Earth as presented in \citeA{Khuller2024}. These measurements were made at several sites:  a glacier in the Purcell Mountains in Canada \cite{Fitzpatrick2017}, oceanic surfaces \cite{Fairall1996, Fairall2003}, an ice-covered lake in the Dry Valleys of Antarctica \cite{Clow1988} and an underground permafrost tunnel in Alaska \cite{Douglas2019}. The same model's input parameters as those used by \citeA{Khuller2024}  are used. The wind used in our model, $U$, is the one provided by measurements and we do not add the gustiness velocity derived in our model \cite{Colaitis2013}. To quantify the validity of our model, we use the same metrics as \citeA{Khuller2024}, i.e., the root mean square error (RMSE), and the dimensionless index of agreement ($d_i$) (see their section 5 for definition). Results are shown in Figure \ref{fig:validation}.

Our model performs reasonably well when modeling the ice sublimation at Purcell Mountains (Figure \ref{fig:validation}a, $d_i$ = 65\% in our model, 69\% for \citeA{Khuller2024}). As in \citeA{Khuller2024}, the main discrepancy between our model and measurements comes from the presence of strong low-level katabatic winds which violates some assumptions of the Monin-Obukhov similarity theory on which our model relies. However, our model does not perform well on smooth ocean surfaces (Figure \ref{fig:validation}b, $d_i$ = 19\% vs 62\% for \citeA{Khuller2024}). The main reasons are: 1) the expression of $z_{0q}$ we used might not be accurate for such smooth surfaces \cite<e.g.,>{Andreas1987}; 2) the absence of gustiness velocity in our model (for this comparison only), while it might play a significant role, especially for low-wind periods ("free-convective periods"). Note that such conditions are unlikely on Mars at gully locations because of the presence of strong slope winds \cite{Smith2018, Spiga2018}.

Our model performs very well at Mars Analog sites on Earth. Figure \ref{fig:validation}c shows that our modeled sublimation rates of ice with the conditions of the dry valleys in Antarctica are in excellent agreement with the ones observed ($d_i$ = 92\%, similarly to \citeA{Khuller2024}). At the permafrost site in Alaska, for wind velocity of 0.05~m~s$^{-1}$ to 0.15~m~s$^{-1}$, and surface roughness between 10 and 15~cm, our model predicts a sublimation rate between 1.6~cm per year to 2.95~cm per year vs 2.1~cm per year measured.  Note that other models, except the one from \citeA{Khuller2024}, predicted a sublimation rate 4 times higher than the one measured \cite{Douglas2019}.

Hence, our model is validated at Mars Analog sites on Earth and glaciers in Canada, but not on the very smooth oceanic surface, especially during free-convective periods. Given that these conditions are unlikely on Mars at gully sites due to the presence of slope winds, we will consider that our model is reliable for this study, although improvements (especially by adding the effect of water buoyancy on the gustiness) will be considered for future studies.

\subsection{Subsurface ice sublimation model}
\label{method:subsurfacesubl}
When ice is buried in the subsurface at depth $z_{\text{ice}}$~(m), and when no frost is present at the surface, the water vapor flux from the subsurface water ice to the atmosphere $F_{\text{ice}}$~(kg~m$^{-2}$~s${-1}$)  is given by (see demonstration in Appendix \ref{appendix:subsurfaceflux}):

\begin{equation} 
     F_{\text{ice}} = 
            \frac{C_q U}{1+\frac{ C_q U z_{\text{ice}}}{D} \times \frac{1}{1-\frac{\rho_{\rm{vap,atm}}}{\rho_{\rm{sv}}(T_{\text{ice}})}}} \left( \rho_{\rm{vap,atm}} - \rho_{\rm{sv}}(T_{\text{ice}}) \right)  
            \label{eq:fluxgroundice}
\end{equation}

\noindent where $D$~(m$^2$~s$^{-1}$), is the diffusion coefficient through the dry material covering the ice, left as a free parameter here. This expression includes the transport of water vapor from the subsurface ice to the surface,  and then the exchange between the surface and the atmosphere with the surface water vapor mixing ratio calculated along the way. The effect of advection within the soil, induced by the gradient of air composition at the subsurface ice interface, is represented by the factor $\frac{1}{1-\frac{\rho_{\rm{vap,atm}}}{\rho_{\rm{sv}}} }$ following \citeA{Hudson2007}. This formulation follows the standard diffusion-limited approach commonly used to describe subsurface ice sublimation in porous media \cite<e.g.,>{Hudson2007,Schorghofer2020}. In case frost (either CO$_2$ or water) is at the surface, the flux is given by:

\begin{equation}
   F_{\text{ice}} =  \frac{D}{z_{\text{ice}}} \frac{1}{1-\frac{\rho_{\rm{vap,atm}}}{\rho_{\rm{sv}}(T_{\text{ice}})} } \left( \rho_{\rm{sv}}(T_{\text{surf}}) - \rho_{\rm{sv}}(T_{\text{ice}}) \right) 
\end{equation}

\subsection{Icy (sub)surface thermal model}

\subsubsection{General description}
The evolution of the ice temperature at the surface is governed by the surface energy budget:

\begin{equation}
  \rho  c_s \frac{\partial T_{\rm{ice}}}{\partial t} = (1-A_{\rm{ice}}) F_{\rm{rad,sw}}(t) + \epsilon_{\rm{ice}}  F_{\rm{rad,lw}}(t) +  F_{\rm{ground}}(t) +  F_{\rm{atm}}(t) + \sum_{i}L_{\text{H}_2\text{O}} F_{\rm{ice}}   - \epsilon_{\rm{ice}} \sigma T_{\rm{ice}}^4(t)
  \label{Eq:energysurflayer}
\end{equation}

\noindent where the left member of the equation is the energy of the ice surface layer, with:
\begin{itemize}
    \item $\rho$  the  density of the ice (kg~m$^{-3}$);
    \item c$_s$  the "surface layer" heat capacity per unit area (J~m~kg$^{-1}$~K$^{-1}$) (heat capacity multiplied by thickness of the first soil layer);
    \item $T_{\rm{ice}}$ the ice temperature (K);
    \item $A_{\rm{ice}}$ (unitless) the albedo of the ice;
    \item $F_{\rm{rad,sw}}$ the radiative flux at visible wavelengths (W~m$^{-2}$);
    \item $\epsilon_{\rm{ice}}$~(unitless)  the surface emissivity (assumed to be equal to the absorptivity);
    \item $F_{\rm{rad,lw}}$ the radiative flux at infrared  wavelengths (W~m$^{-2}$);
    \item  $F_{\rm{ground}}$ the soil heat flux due to heat conduction process (W~m$^{-2}$);
    \item $F_{\rm{atm}}$ the sensible heat flux (W~m$^{-2}$);
    \item $L_{\text{H}_2\text{O}} F_{\rm{ice}}$ the sublimation cooling, with $L_{\text{H}_2\text{O}}$~=~2830 J~kg$^{-1}$ is the latent heat of sublimation of water;
    \item $ \epsilon_{\rm{ice}} \sigma T_{\rm{ice}}^4$ the radiative cooling of the surface with $\sigma$~=~5.67$\times 10^{-8}$ W~m$^{-2}$~K$^{-4}$ the Stefan-Boltzmann constant. 
\end{itemize}
\noindent A detailed description of all the terms can be found in \cite{Lange2023Model}.
By default, the emissivity of ice is set to 1, and a broadband albedo of water ice of 0.33 is used. Although this value is quite low for pure H$_2$O frost, unless it is either mixed with dust \cite[for dust concentrations below ~1\%]{Khuller2021albedo} or optically thin \cite[less than a few millimeters thick]{Singh2016}, it allows us to accurately reproduce the present-day Martian seasonal water cycle \cite{Navarro2014} and the seasonal evolution of water frost on slopes \cite{Lange2023Model}, and remains consistent with observed water frost albedo values, which typically range between 0.2 and 0.6 \cite{Langevin2005}. Varying the albedo used in the simulations do not change the conclusion drawn in this manuscript.

When ice is present in the subsurface, the computation of it temperature is based on a 1D implicit soil model and solves the heat soil conduction:

\begin{equation}
    \rho C_p \frac{\partial T_{\text{soil}}}{ \partial t} = \frac{\partial}{\partial z} \left(\lambda(z) \frac{\partial T_{\text{soil}}}{\partial z}  + L_{\text{H}_2\text{O}} \times F_{\text{ice}} \right)
    \label{eq:soiltemp}
\end{equation}

\noindent where $\rho$~(kg~m$^{-3}$) is the material's density, and $C_p$~(J~kg$^{-1}$~K$^{-1}$) is its specific heat, $T_{\text{soil}}$~(K) is the subsurface temperature, $z$~(m) is the depth, $\lambda$~(W~m$^{-1}$~K$^{-1}$) is the thermal conductivity. The soil grid ranges from 0.2~mm to $\sim$~20~m, with an irregular grid with a fine resolution ($\le$~mm) close to the surface. When subsurface ice is present at a depth $z_{\text{ice}}$~(m), the thermal inertia (and thereby conductivity) of the corresponding layer and the layers underneath are set to a value $I_{\text{ice}}$~(\tiu). When ice is condensing/subliming, a latent heat flux $L_{\text{H}_2\text{O}} \times F_{\text{ice}}$ is added to the energy budget. By default, we set $I_{\text{ice}}$~=~1600~\tiu, a mid-value between completely pore-filled ice \cite<thermal inertia of $\sim$1200~W~m$^{-2}$~K$^{-1}$; Eq. (26) of >{Siegler2012} and massive pure ice \cite<thermal inertia of $\sim$2050~W~m$^{-2}$~K$^{-1}$ at 180~K;>{Hobbs1974}. This value allows for a realistic representation of the present-day Martian CO$_2$ cycle, which is known to be sensitive to this effect, as well as the distribution of frost on pole-facing slopes \cite{Lange2023ice}.

\subsubsection{Model limitations}
\label{ssec:modellimits}
Although the climate simulations using the model presented above agree with the observed seasonal and spatial distribution of water frost on Mars \cite{Vincendon2010water, Navarro2014,Lange2023Model}, their surface temperatures \cite{Lange2024}, and the presence of subsurface water ice at high latitudes \cite{Lange2023ice}, the model neglects the penetration of solar radiation into the ice. As shown in the previous section, solar radiation is assumed to be absorbed only at the surface, that is, within the first millimeters of the ice, and not transmitted to deeper layers.

This assumption is consistent with freshly deposited snow mixed with surface dust, which renders it opaque, or with ice covered by a surface dust lag. Near-infrared observations indicate that high-latitude water frost on Mars has typical grain sizes of 10–100~$\mu$m \cite{Langevin2005,Langevin2007} supporting the use of this assumption for seasonal deposits. The latter situation is expected to occur primarily during the sublimation of dusty ice, since dust sintering tends to clean the remaining ice. Our calculations are thus valid only under this assumption, consistent with the simulations presented in section~\ref{sec:surfacemelt}, which explore the melting of frost or snow deposited during winter. Other simulations consider ice covered by a few millimeters of dust, which should effectively block solar radiation.

As snow undergoes metamorphism, the grain size is expected to increase, allowing solar flux to penetrate deeper and potentially warming the ice enough to trigger localized melting (see section \ref{discussion:meltdustyice}). However, the metamorphism rate remains poorly constrained, with estimates ranging from 10–100 years \cite{Clow1987} to more than one million years \cite{Bramson2017}, making it difficult to assess whether this process can operate on seasonal timescales. If, nevertheless, coarse-grained snow or relatively clean ice are exposed at the surface, this effect could become significant and potentially allow melting (see discussion in section \ref{discussion:meltdustyice}).

\section{Results}
\subsection{Melting frost on the surface}
 \label{sec:surfacemelt}
\begin{figure}[h!]
  \centering
  \includegraphics[width = \textwidth]{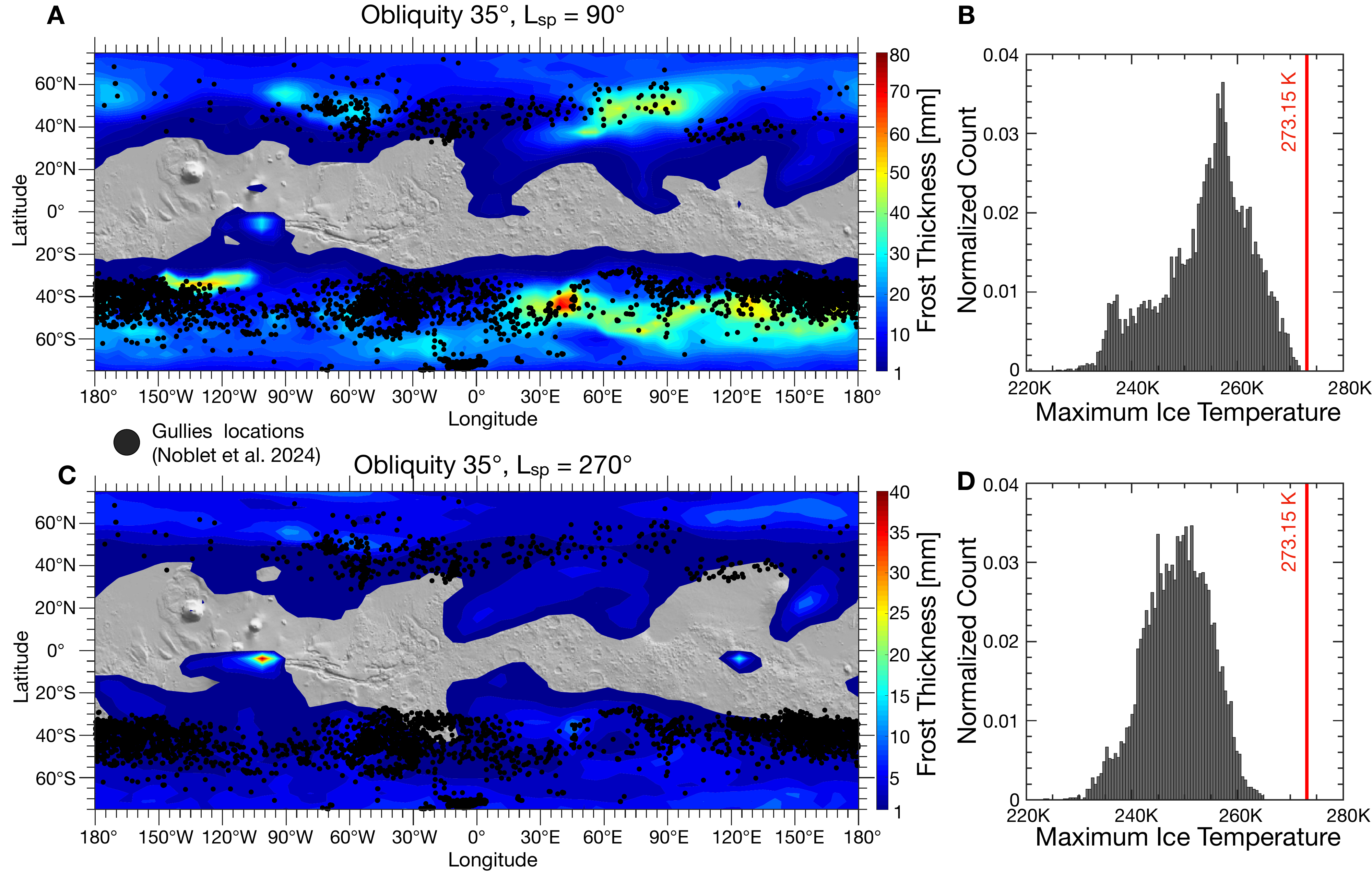}
  \caption{\textit{Left column}: Maximum thickness of seasonal frost on 30\textdegree~pole-facing slopes for an obliquity of 35\textdegree~and \Lsp~=~90\textdegree~(a), and  \Lsp~=~270\textdegree~(c) Gullies locations from \citeA{Noblet2024} are reported by black dots. \textit{Right column}: Maximum  temperatures of seasonal frost on 30\textdegree~pole-facing slopes for an obliquity of 35\textdegree~and \Lsp~=~90\textdegree~(b), and  \Lsp~=~270\textdegree~(d). The red line is the melting point of water ice (273.15~K).}
  \label{fig:melting_frost}
\end{figure}

We first test the possibility of melting opaque snow/frost on the surface. As theoretically shown by \citeA{Ingersoll1970}, \citeA{Schorghofer2020} and \citeA{Khuller2024}, melting of frost/ice on present-day Mars is unlikely because of the cooling of the ice by the latent heat released during the sublimation. \citeA{Lange2024}  showed that frost temperatures measured on Mars and predicted by the PCM never reach the melting point of water ice (see their Figure 10). As explained in \citeA{Lange2024}, the low humidity of the present-day Martian atmosphere limits frost thickness to at most a few millimeters and leads to strong sublimation rates, and thus strong sublimation cooling, preventing frost from reaching high temperatures. To maximize the possibility of surface frost or ice melting, both atmospheric humidity and the solar flux incident on sloped surfaces must be maximized. These conditions occur during the last 4 million years when obliquity reaches ~35\textdegree and perihelion coincides with northern summer (\Lsp~=~90\textdegree), which maximizes solar insolation on the northern polar water ice cap. We therefore perform climate simulations and compute water frost thickness and temperature under these optimal conditions (Figure \ref{fig:melting_frost}). For completeness, we also consider the opposite configuration (obliquity 35\textdegree, \Lsp~=~270\textdegree). We present below results for 30\textdegree~pole-facing slopes, as they favor frost formation during winter due to low insolation and maximize frost and ice temperatures during spring and summer by receiving higher insolation than flatter surfaces \cite{Costard2002,Lange2024}. At high obliquity, when the atmosphere is most humid over the last 4 million years, these slopes therefore represent the most favorable configuration for surface melting.

For an obliquity of 35\textdegree, water frost forms down to $\sim$~20\textdegree~latitude on 30\textdegree~pole-facing slopes in both hemispheres and closer to the Equator locally in low-thermal inertia regions. The frost thicknesses are larger for \Lsp~=~90\textdegree~than 270\textdegree~as a consequence of the difference in global atmospheric humidity between the two simulations. Indeed, for the simulation with  \Lsp~=~90\textdegree, the Northern glacier sublimes more, giving an average humidity value of 260 pr-\mum, whereas it is only  58 pr-\mum~for \Lsp~=~270\textdegree. Nevertheless, in both cases, the frost thicknesses are much ($\sim$~30 times for \Lsp~=~90\textdegree) higher than observed on present-day Mars. As water frost is thicker, it remains longer on the surface and can warm up more as it is more exposed to the sun. In addition, the sublimation flux, and thereby sublimation cooling, is reduced because of the higher near-surface atmospheric humidity. However, as shown in Figures \ref{fig:melting_frost}a, c, water frost can neither reach the melting temperature, again because of the sublimation cooling by latent heat. The maximum frost temperatures are 271.8~K for \Lsp~=~90\textdegree, and 264.9~K for \Lsp~=~270\textdegree. Using the 1D model, we test the effect of the ice properties on the maximum frost temperatures. Even using a low broadband albedo (down to 0.2, representing very dusty ice), the frost temperatures can never reach 273.15~K and melt. Note that using a higher albedo decreases the probability of melting, as frost and ice reflect more solar radiation and absorb less energy. Hence, we confirm that the melting of water frost/  ice on the surface is unlikely because of the sublimation cooling, even at high obliquity when the planet is much wetter, refuting the conclusions of \citeA{Kossacki2004} and \citeA{Dickson2023} which did not include this effect. The possibility of lowering the melting temperatures in the presence of salts and of melting within the ice at depth due to a solid-state greenhouse effect are discussed in section \ref{sec:discussmelt}.

\subsection{Melting subsurface water ice in equilibrium with the atmosphere}
\label{sec:stableicetemp}

\begin{figure}[h!]
  \centering
  \includegraphics[width = \textwidth]{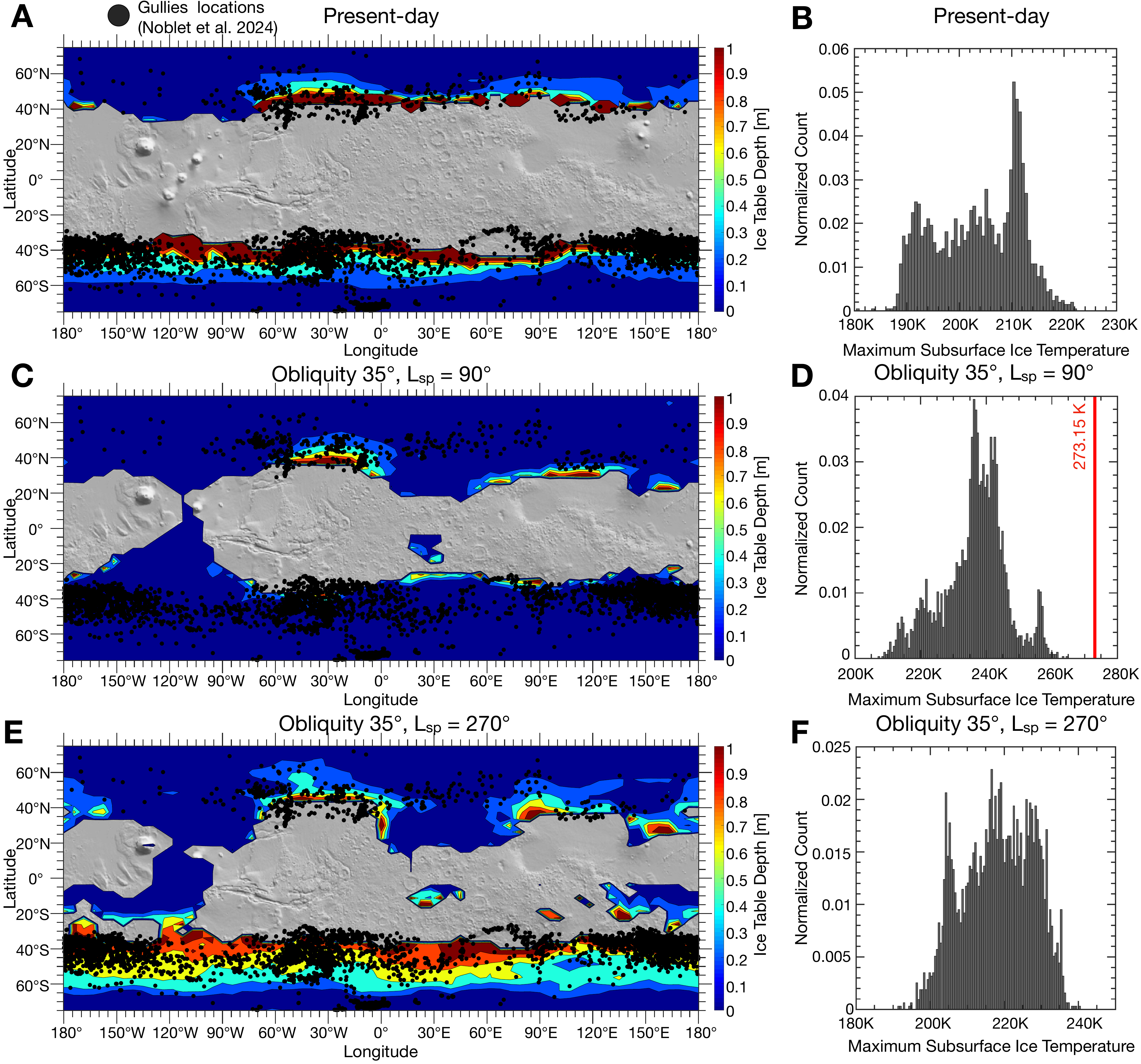}
  \caption{\textit{Left column}: Depth at which ice table beneath a 30\textdegree~pole-facing slope for present-day (a),   an obliquity of 35\textdegree~and \Lsp~=~90\textdegree~(c), and  \Lsp~=~270\textdegree~(e). Gullies' location from \citeA{Noblet2024} is reported by black dots. \textit{Right column}: Maximum temperatures of the upper surface of the ice table for present-day (b), an obliquity of 35\textdegree~and \Lsp~=~90\textdegree~(d), and  \Lsp~=~270\textdegree~(f). The red line is the melting point of water ice (273.15~K).}
  \label{fig:melting_ssi_stable}
\end{figure}

 \citeA{Costard2002} proposed that subsurface ice present even below the diurnal thermal skin depth could melt as diurnal mean temperatures on 30\textdegree~pole-facing slopes at mid-latitudes exceed 273.15~K at high obliquity. However, \citeA{Mellon2001} reported that such ice should disappear completely before reaching the melting temperature. To test this possibility, we first compute the depth at which the subsurface is in equilibrium with the water vapor content in the atmosphere for present-day and 35\textdegree~obliquity with the 3D model. Subsurface ice is  in equilibrium with the water vapor content in the atmosphere at depth $z_{\rm{ice}}$~(m) when the yearly-averaged water vapor density at the surface is equal to the one at the ice table \cite{Mellon2004, Schorghofer2005}. The thermal inertia of the layer below this depth is set to 1600~\tiu. Simulations are relaunched to include the effect of this large thermal inertia on the soil temperatures, and the process is iterated until the equilibrium depth has converged. Equilibrium depths for present-day climate and 35\textdegree~obliquity are presented in Figure \ref{fig:melting_ssi_stable}.

The distribution of subsurface ice on pole-facing slopes for the present climate is discussed in \citeA{Lange2023ice}.  At an obliquity of 35\textdegree~and \Lsp~=~90\textdegree, the modeled atmospheric humidity is so high that ice remains is in equilibrium with the water vapor content in the atmosphere within the upper centimeters of the surface (Figure \ref{fig:melting_ssi_stable}c)  at tropical latitudes and locally in dusty equatorial regions. These depths do not necessarily imply the presence of massive ice deposits at these latitudes and depths: they rather indicate that if such ice exists, it is is in equilibrium with the water vapor content in the atmosphere. However, even if massive ice is not present, ice can still form within the pores of the regolith. At an obliquity of 35\textdegree~and \Lsp~=~270\textdegree, ice is not in equilibrium with the water vapor content in the atmosphere  near the surface, but is in equilibrium  at greater depths ($>$~70~cm) due to lower atmospheric humidity.

In all cases, we find that subsurface ice does not reach the melting temperature: the maximum temperature is 259~K, which occurs at an obliquity of  35\textdegree~and \Lsp~=~90\textdegree, when the ice is closest to the surface (Figures \ref{fig:melting_ssi_stable}c, d). For the present-day, the maximum subsurface ice temperature is 222.4~K and is 240.5~K at obliquity of  35\textdegree, \Lsp~=~270\textdegree.  This result is expected: when stable, ice is generally not exposed to large diurnal or seasonal temperature variations; otherwise, sublimation would be too significant, preventing it from remaining in its equilibrium zone.  Changing the thermal inertia of subsurface ice from 1600 to 800~\tiu~for instance does not change our conclusions: if a lower thermal inertia is used, the ice buries itself deeper than it would with a higher thermal inertia. This feedback mechanism ensures that the ice remains cold on average throughout the year. Finally, we have checked whether pore-filling ice or subsurface frost can melt. In all cases, the quantity of frost is too low and it sublimes very quickly before reaching the melting temperature of ice. We conclude that when ice is at its equilibrium depth, it cannot melt.

\subsection{Melting suddenly warmed subsurface water ice}
\label{ssec:meltunstable}
As shown previously,  ice at the surface can not melt because of the latent heat cooling induced by the significant sublimation rate. Ice buried beneath the regolith can be protected from sublimation and thereby latent heat cooling, but ground ice in equilibrium with the atmosphere is too deep to be warmed enough and melt. Can we imagine a process that would allow stable ice on Mars to be exposed to an unusual or even sudden heating? For instance, a well-known case on Mars is when a meteoritic impact in the mid-latitude excaves regolith and sometime exposes water ice \cite{Byrne2009}, but such events are rare and local. However, another process that can play this role exist on Mars and it is precisely active on gullied slopes: there, at the end of winter, seasonal CO$_2$ ice sublimation has been shown to sometime destabilizes the surface regolith material \cite{Dundas2012, Pilorget16}, inducing a dry debris flow, that is now though to be at the origin of most gully activities on Mars today. When such a process occur, the local water ice table (almost always present on known gullied slopes)  can suddenly be exposed at the surface \cite{Khuller2021ssice} or left with only a thin layer of dry soil above it \cite{Pilorget16}, experiencing rapid heating when the surface is heated by the sun but without losing too much latent heat thanks to the thin soil cover between the water ice and the atmosphere.  We also note that, in this scenario, water frost trapped at the base of the seasonal CO$_2$ ice layer could in principle experience enhanced springtime heating, as solar radiation can penetrate the translucent CO$_2$ ice. However, as shown by \citeA{Pilorget2011} and \citeA{Pilorget16}, the temperatures reached at the CO$_2$ ice–soil interface, where such water frost would be located, remain below $\sim$160~K, well below the melting point of water, and therefore do not allow liquid water formation.

\begin{figure}[h!]
  \centering
  \includegraphics[width = \textwidth]{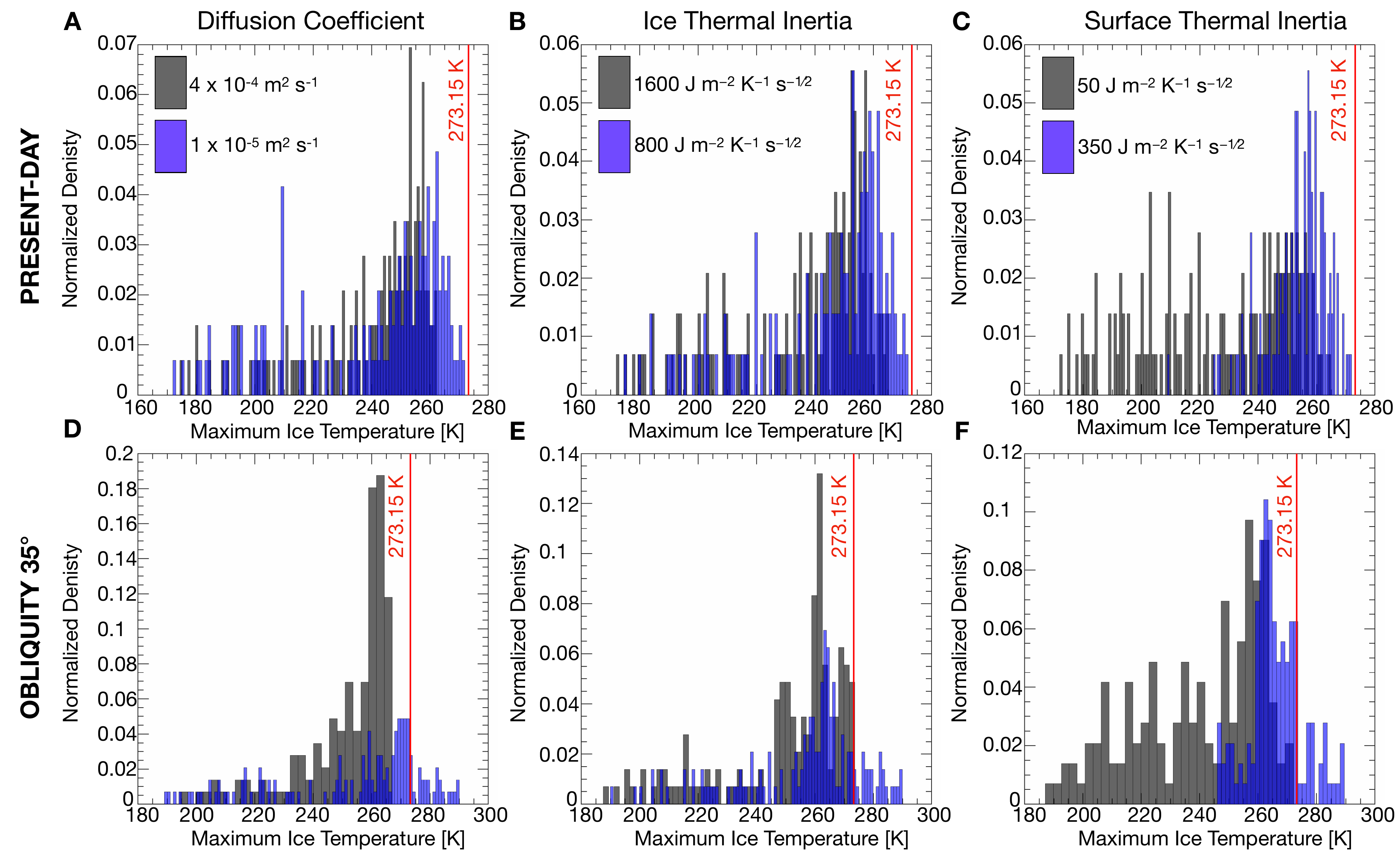}
  \caption{Distribution of maximum subsurface ice temperatures simulated after surface erosion. Each count represents a simulation with parameters detailed in Table \ref{Table:inputs}, normalized by the total number of simulations (288). Top panels (a–c) correspond to present-day obliquity and water vapor content, while bottom panels (d–f) correspond to 35\textdegree~obliquity. Panels a and c illustrate sensitivity to the dry soil diffusion coefficient; b and e to ice thermal inertia; c and f to dry soil thermal inertia. Only ice temperatures beneath 30\textdegree~slopes are shown, as this configuration produces the highest ice temperatures during summer. Temperatures may exceed the melting point because the latent heat of melting is not included in our model once ice temperatures exceed 273.15~K.}
  \label{fig:meltingexp}
\end{figure}

To test whether subsurface ice can melt under the latter conditions, we use our 1D model to simulate its evolution after surface erosion (inputs in Table \ref{Table:inputs}). Subsurface ice is initially placed at depth $z_{\text{ice}}$, and the layer below are assigned ice thermal properties. The model equilibrates temperatures over 10 years. After that, during the  CO$_2$ sublimation in winter, ice is moved from $z{\text{ice}}$ to $(1-\alpha) z_{\text{ice}}$, with $\alpha$ being an excavation factor between 0 (no excavation) and 1 (ice exposed at the surface). The removed soil thickness $\alpha z_{\text{ice}}$ defines the new surface, setting the soil temperature there. The model then calculates sublimation, temperature, and ice depth evolution over 2 years (1‑minute timestep). We will only consider here the case in which ice is still opaque as it its covered by a thin layer of dust. Indeed, our model might not be appropriate to study the case of suddenly exposed ice, when solar absorption could have a significant impact on the possibility of melting (see section \ref{discussion:meltdustyice}). Results of our experience are shown in Figure \ref{fig:meltingexp}.

 \begin{table}[h]
 \begin{center}
     
 \begin{tabular}{p{0.45\textwidth}p{0.65\textwidth}} \hline
Parameter & Value/Range of values  \\
\hline
Obliquity	& 25.2\textdegree, 35\textdegree  \\
Solar Longitude of Perihelion	& 270\textdegree  \\
Eccentricity & 0.0934 \\
Latitude & 30\textdegree S, 40\textdegree S, 50 \textdegree S \\
Slope angle & 0\textdegree-30\textdegree, 10\textdegree~step, facing North and South \\ 
Subsurface ice initial depth $z_{\rm{ice}}$ & 0.2 – 1~m\\
Albedo of bare ground & 0.13 \\
Thermal inertia of dry soil & 50 - 350~\tiu \\
Surface pressure (present-day) & 400 - 800 Pa \\
Vapor in the atmosphere (present-day) & 10 pr-$\mu$m\\
Surface pressure (35\textdegree~obliquity) & 800 - 1600 Pa \\
Vapor in the atmosphere (35\textdegree~obliquity & 260 pr-$\mu$m\\
Diffusion coefficient & 4~$\times$~10$^{-4}$, 1~$\times$~10$^{-5}$~m$^2$~s$^{-1}$ \\
Excavation factor $\alpha$ & 0.9, 0.99, 0.999 \\
\hline
 \end{tabular}
 
 \end{center}
\caption{ Set-up for the simulations performed in section \ref{ssec:meltunstable}. Simulations are done in the South as it is where most of the gullies are observed.}
\label{Table:inputs}
 \end{table}

\subsubsection{Present-day:} 
For current orbital conditions, the temperature of the subsurface ice after being destabilized never reaches 273.15~K. The median temperature reached is 248~K, and the maximum temperature is 271.8~K. We find that the maximum temperature achieved depends on surface and subsurface properties:

\begin{itemize}

    \item Ice must be close to the surface, otherwise it will be too deep to be heated by the diurnal/seasonal thermal heat wave during summer. However, the ice cannot be at too shallow a depth otherwise sublimation cooling would reduce the ice temperature. As shown analytically by \citeA{Schorghofer2020} and confirmed numerically by our simulations, ice located a few centimeters below the surface represents an optimum configuration that maximizes heating while minimizing latent heat cooling.
    
    \item The ice temperature is higher when the covering dry soil diffusion coefficient is low (Figure \ref{fig:meltingexp}a). If the diffusion coefficient is too high, then the sublimation of the ice is strong and the sublimation cooling prevents reaching high ice temperatures. According to \citeA{Hudson2007}, the possible range of diffusion coefficients on present-day Mars is between $10^{-5}$~m$^{2}$~s$^{-1}$ (compacted dust) and $4 \times 10^{-4}$~m$^{2}$~s$^{-1}$ (regolith). In comparison, effective diffusion coefficient of the ice-free permafrost on Earth is about $10^{-6}$~m$^{2}$~s$^{-1}$ \cite{McKay1998,Douglas2019}. Within the range of typical Martian diffusion coefficients, our model does not predict ice melting (Figure \ref{fig:meltingexp}a).

   \item  The thermal inertia of the ice must be low. This point will be discussed below, but even when using unrealistically low thermal inertia values (e.g., $<$~100~\tiu), our model does not predict the melting of the ice (Figure \ref{fig:meltingexp}b).

   \item 
   The thermal inertia of dry soil, and thus its conductivity, must be high. More conductive soil allows heat to penetrate deeper, reaching buried ice. However, higher conductivity also reduces the diurnal temperature peak, lowering the likelihood of melting. In any case, no melting occurs in our model (Figure \ref{fig:meltingexp}c). Note that higher conductivity requires cementation between soil constituents \cite{Piqueux2009a, Piqueux2009b}, which reduces the material's diffusion coefficient, making this condition consistent with the first.

\end{itemize}

\subsubsection{Recent paleoclimates:} During periods of high obliquity, in a few extreme simulations, the modeled maximum subsurface ice temperature (without including the latent heat of fusion) can be higher than 273.15~K (Figure \ref{fig:meltingexp}), meaning that ice could melt. The melting occurs in the afternoon during summer, over a few days. We estimate (see method in Text S4) that the amount of water formed does not exceed 0.5~mm, i.e., 5\% of the volume of the regolith layer above ice (Figure S4). Such value is below the 10\% threshold concentration required for debris flow formation \cite{Malin2000, Costard2002}. Moreover, an analysis of the combined parameters requested to reach melting suggest that they are not realistic. First, melting only occurs if the subsurface diffusion coefficient is less than $10^{-5}$~m$^2$~s$^{-1}$ (the maximum temperature of ice covered by a surface with a diffusion coefficient of $4 \times 10^{-4}$~m$^2$s$^{-1}$ is 267~K). Such a low diffusion coefficient is difficult to imagine on Mars due to the low atmospheric pressure. Even assuming a doubling of pressure during periods of high obliquity, the diffusion coefficient remains larger than $4-5 \times 10^{-5}$~m$^2$~s$^{-1}$ for regolith or dust with micrometer-sized grains \cite{Hudson2007, Schorghofer2020}. Our model does not predict the melting of the ice when using these values  (not shown).  \citeA{Hudson2008} experimentally explored the possibility of reducing the soil diffusion coefficient on Mars. Specifically, they found that only mechanically compacted dust, after being buried deeply in the subsurface, may effectively reduce the diffusion coefficient, potentially reaching values as low as $10^{-5}$~m$^2$~s$^{-1}$. Therefore, only ancient ice, initially buried at depth before surface erosion, might be likely to melt.

When using such a low diffusion coefficient, our model shows that only ice with thermal inertia of $\leq 800$~\tiu~can melt. By varying the thermal inertia of the ice, we observed that a few simulations show melting at 1000~\tiu, but this remains rare. No simulations predict melting for ice with a higher thermal inertia. Such a low thermal inertia for ice is associated with either:

\begin{itemize}
    \item Snow undergoing metamorphism to ice \cite<thermal inertia values of $\leq$~1000~\tiu~are associated with an ice/snow porosity of 50\%, see for instance  Figure 3 of>{Bapst2019}. Although the rate of ice metamorphism is not well constrained \cite<see discussion in>{Khuller2021ssice}, this scenario seems unrealistic. As previously shown, the ice must initially be deeply buried. \citeA{Bramson2017} showed that mid-latitude subsurface ice buried beneath a few meters of regolith has a porosity of $\sim$~20\%, translating into a high thermal inertia of more than 1600~\tiu \cite{Bapst2019}. It is therefore reasonable to assume that ice buried beneath slopes would undergo a similar degree of metamorphism, and hence exhibit a comparably high thermal inertia.
    \item Ice filling the pores of the regolith \cite{Siegler2012}. However, this scenario is also unrealistic because the material above the ice would need to be compacted, making it difficult to imagine ice in the pores.
\end{itemize}

For realistic conditions of dry soil, ice thermal inertia and diffusion coefficient, the maximum temperature obtained is below 273~K. Four simulations are within 1~K of the melting temperature of ice, and the rest are colder. Only the simulations with unrealistic properties (thermal inertia lower than 800~\tiu, diffusion coefficient lower than $10^{-5}$~m$^2$~s$^{-1}$) leads to the melting of the ground ice.  When changing some parameters (e.g., the albedo of the surface, depth of the ice after the surface removal, or eccentricity of the planet), a few simulations (less than 5\% of the total number of simulations performed)  manage to reach the melting temperature of water ice. Hence, we conclude that the melting of subsurface ice after being destabilized is very unlikely in the recent past of Mars.

\subsubsection{Melting during the formation of a lag-deposit:}
Another possibility to have unstable shallow subsurface ice could be during the formation of a lag deposit associated with with the sublimation of a dusty ice deposit.
As dusty ice sublimes, dust in the ice accumulates on the surface. This process is thought to create lag deposits above the ice, composed of uncompressed (high diffusion coefficient) micron-sized dust particles (with a low thermal inertia). We tested whether melting is possible in this case by modeling the ice temperature beneath a layer of dust (thickness ranging from 1~mm to 1~m, thermal inertia of 50~\tiu, other properties are set as given in Table \ref{Table:inputs}) for present-day and past obliquity. In all of our simulations, the melting temperature is never reached (maximum temperature is 266~K). Indeed, as dust poorly conducts heat, ice must be exposed close to the surface to be effectively warmed. However, because dust has a high diffusion coefficient \cite<down to 4$\times$10$^{-5}$~m$^2$~s$^{-1}$,>{Hudson2007, Hudson2008, Schorghofer2020}, it does not reduce significantly the sublimation of shallow ice and thereby latent heat cooling. Consequently, ice is very unlikely to melt in this case.

\section{Discussion}
\label{sec:discussmelt}
\subsection{Effects of salts on the melting of water ice}
 The presence of salts,  widespread on the Martian surface \cite<e.g.,>{Clark1981, Hecht2009, Glavin2013}, in the water ice mixture can reduce the temperature needed to melt. For most of our simulations, frost at the surface, or ice buried in the surface (at equilibrium or unstable) typically have a temperature 20-30~K lower than the melting temperature of pure water ice. The concentration of salt required to reduce the freezing point to such low temperature is significant,  up to 40\% by weight \cite<Figure 7 of>{Mellon2001, Toner2015}.  These values are significantly higher than the amount of salts measured in the martian regolith. For example, the abundance of perchlorates measured at various landing sites is $<$~1~wt.\% \cite<see Table 6 in>{Glavin2013}. Cemented surfaces, such as duricrusts, can contain up to a few percent of salt \cite<e.g.,>{Cabrol2006}, and one can reasonably assume that this amount should be much less within ice deposits formed by snowfall.  At such low abundances, the melting temperature is almost similar to that of pure water ice \cite{Toner2015}. Therefore, while localized melting at the salt–ice interface may occur, the formation of substantial brine volumes, and hence their influence on gully activity, appears to be limited or unlikely. In a few cases, when the temperature of the ice is within 1-2~K the melting temperature of pure ice, salts concentrations might be high enough to depress the melting temperature and allow the formation of larger brines. Yet, as these cases are rare, we conclude that the formation of large volume brines through the melting of salty ice is unlikely in the recent past of Mars. We acknowledge that a fully self-consistent treatment of the effects of salts on ice sublimation and melting would require explicitly accounting for their impact on thermodynamic properties and mass transfer, including melting temperatures, equilibrium vapor pressures, sublimation fluxes, as well as potential deliquescence \cite<e.g.,>{Chevrier2022} and changes in the effective diffusion coefficient \cite<e.g.,>{Hudson2008} of the porous medium. Such complexity is beyond the scope of the present study and is left for future work. We also acknowledge that highly concentrated salt solutions could form on Mars through deliquescence \cite<e.g.,>{RiveraValentin2020} or repeated thin-film wetting \cite<e.g.,>{Mohlmann2008}. However, the resulting brines are expected to occur only in very small amounts \cite{Chevrier2024} and are very unlikely to generate large debris flows such as gullies.

\subsection{Melting ice through the solid-state greenhouse effect}
\label{discussion:meltdustyice}

As mentioned in section \ref{ssec:modellimits}, simulations performed in this study assume that water ice/snow is opaque, meaning all incoming solar radiation is absorbed at the surface of the ice and does not penetrate the ice. In reality, ice can be transparent, allowing photons to heat it from within. If the ice contains impurities such as dust, these impurities can absorb downwelling photons, leading to a solid-state greenhouse effect \cite{Clow1987, Williams2009, Khuller2024absorption} and heating at depth. In addition, the overlying snowpack is close to saturation. As a result, sublimation and evaporation at depth are reduced, thereby limiting latent heat losses and further promoting melting and the persistence of liquid water. \citeA{Clow1987, Williams2008} and \citeA{Williams2009} showed that even a small amount of dust ($\le$~1000 parts per million by weight) within the ice is enough to absorb sufficient solar radiation to generate a solid-state greenhouse effect strong enough to melt snow or ice at depth. Notably, \citeA{Williams2009} demonstrated that melting could occur if ice with a small amount of dust is exposed at the surface in late spring or summer, when solar heating is at its maximum, producing approximately 1~L~m$^{-2}$ of liquid water per day.

Water ice on Mars can contain impurities such as dust. While ice in the polar layered deposits is inferred to contain relatively high dust fractions \cite<more than 10\% of the ice content,>{Plaut2007,Zuber2007,Lalich2019}, observations of mid-latitude water ice at impact crater sites where subsurface ice has been excavated \cite{Byrne2009, Dundas2018, Dundas2021ice}, at scarps \cite{Dundas2018}, and in the alcoves of some gullies \cite{Khuller2021ssice} suggest that this ice is relatively clean, with dust contents below 1\%. The melting of such ice deposits might occur through the mechanism proposed by \citeA{Clow1987, Williams2008} and \citeA{Williams2009}. However, since these observations are limited to specific sites rather than all active gully locations \cite{Rangarajan2024}, we cannot draw quantitative conclusions about how frequently this process may contribute to gully activity, nor about the prevalence of melting driven by this mechanism in Mars' recent past. Moreover, the potential for melting also depends on ice microphysics, notably the grain size (as larger grains enhance deeper solar absorption) and the dust content within the ice. Given the limited constraints on Martian ice metamorphism and the past dust cycle \cite{Kahre2017}, we are therefore unable to estimate the frequency of melting episodes resulting from the solid-state greenhouse effect in Mars' recent history, which remains the subject of ongoing investigation \cite<e.g.,>{Khuller2024absorption}.

\subsection{Other mechanisms}
Other mechanisms have been proposed to promote ice melting. The first concerns the radiative effect induced by the geometry of gully alcoves. These alcoves create shading that forms cold traps where volatiles, such as CO$_2$ and water, preferentially condense during winter. When exposed to sunlight, these points can become warm due to direct insolation, limited radiative cooling to the sky by surrounding walls, and radiative fluxes emitted by these walls. \citeA{Hecht2002} suggested that these effects, combined with a clear sky and a poorly conductive surface, could allow ice to melt in these locations. However, the effect strongly depends on geometry, particularly the depth-to-width ratio of the alcove, which controls the balance between maximizing insolation and minimizing radiative losses. \citeA{Schorghofer2019}, using a more complete model including shadowing and radiative coupling, showed that while these alcoves promote preferential condensation of water frost, the frost disappears before melting occurs. Since these simulations were performed for a single crater (Palikir Crater), new simulations at active gully sites would be required to quantitatively assess the influence of these geometric effects on gully activity.

A second possibility is melting of subsurface ice due to geothermal heating. \citeA{Mellon2001} showed that a combination of typical geothermal fluxes and a poorly conducting soil layer of ~100 m thickness could allow the formation of subsurface aquifers. \citeA{Wood2007} further suggested that pressure variations linked to orbital cycles \cite{Buhler2021} could reduce the thermal conductivity of the regolith, leading to deeper warming of subsurface ice through geothermal heat trapping, thus promoting melting as proposed by \citeA{Mellon2001}. However, as noted by \citeA{Mellon2001}, this mechanism relies on the presence of a thick, poorly conducting regolith layer. Although the thickness of the Martian regolith is poorly constrained, observations at the InSight landing site show rocks a few meters below the surface \cite{Golombek2020}, which contradicts the assumption of a thick regolith layer, although this may not be representative of the entire planet. Furthermore, no observations currently support the existence of subsurface aquifers, rendering this mechanism uncertain.

The last mechanism involves the excavation of subsurface ice by impacts \cite{Byrne2009}. The energy delivered by the impactor could be sufficient to melt buried ice locally. Modeling by \citeA{Reufer2010}, based on the impacts detected by \citeA{Byrne2009}, indicates that melting only occurs in the largest craters, where the impactor penetrates deeply into the ground ice (at least $\sim$1~m). However, no mud, cemented material, or flow-like features generated by this liquid water have been observed \cite{Dundas2014}, and due to the spatial and temporal rarity of such impacts, liquid water formation via this mechanism would remain a rare phenomenon.
\section{Conclusions}
\label{sec:conclusionsmelt}
The objectives of this paper were to discuss the possibility of melting opaque water ice/snow at the surface or in the subsurface of Mars during the last 4 million years when the obliquity of the planet was lower than 35\textdegree. The main conclusions of this investigation are: 
\begin{enumerate}
    \item Fresh deposited water frost/snow at the surface cannot melt on present-day Mars and during recent periods of high obliquity (up to 35\textdegree) because of the significant sublimation cooling which prevents ice from reaching 273.15~K. 
    \item Subsurface water ice at its equilibrium depth can not melt as it is too deep to be efficiently warmed.
    \item Subsurface ice destabilized after an erosion of the surficial regolith (induced by CO$_2$ ice sublimation for instance) is unlikely to melt since melting requires unrealistic conditions (porous ice that has only been half metamorphised, while being deeply buried and thus old).
    \item Melting during the burial of the ice is also unlikely as the regolith/dust can not act as an efficient diffusion barrier. 
    \item The presence of salts could allow the formation of brines at the salt–ice interface. However, the volume of the liquid brines would be low given the low concentration of salt that does not significantly depress the melting temperature of ice.
     \item Exposure of ice with a small amount of dust could melt because of the solid-state greenhouse effect induced by the absorption of solar radiation by dust \cite{Clow1987, Williams2008, Williams2009}. However, we can not draw quantitative conclusions about the occurrence of such a mechanism in the recent past of Mars given the sparse observations of such ice at gully locations \cite{Khuller2021ssice, Rangarajan2024}.
\end{enumerate}

Although our study focuses on the recent past of Mars, it cannot be generalized to earlier times, particularly in periods of very high obliquity ($\sim$45\textdegree), when the planet was even more humid, cloudy and likely dustier, possibly allowing ice to melt. Finally, improved constraints on Martian ice metamorphism, including the evolution of grain size, porosity, and dust content over time, are essential to better assess the potential for melting in both recent and earlier climatic periods. Such knowledge would refine our understanding of the conditions under which surface and subsurface ice could transiently produce liquid water on Mars.

\section*{Conflict of Interest}
The authors declare no conflicts of interest relevant to this study.
\section*{Open Research}
\begin{itemize}
    \item Data: Data files for figures used in this analysis are available in a public repository, see \citeA{Lange2025Marsdata}. 
    \item  Software: The Mars PCM  used in this work can be downloaded with documentation at \url{https://svn.lmd.jussieu.fr/Planeto/trunk/LMDZ.MARS/.}.  More information and documentation are available at http://www-planets.lmd.jussieu.fr.   \cite{Forget2025PCM}
\end{itemize}

\acknowledgments
This project has received funding from the European Research Council (ERC) under the European Union's Horizon 2020 research and innovation program (grant agreement No 835275, project "Mars Through Time"). Mars PCM simulations were done thanks to the High-Performance Computing (HPC) resources of Centre Informatique National de l'Enseignement Supérieur (CINES) under the allocation n\textdegree A0100110391 made by Grand Equipement National de Calcul Intensif (GENCI);  to the IPSL Data and Computing Center ESPRI which is supported by CNRS, Sorbonne Université, CNES and Ecole Polytechnique, and to the SACADO MeSU platform at Sorbonne Université. The authors acknowledge A.Khuller, N.Schorghofer, C.Freissinet, V.Chevrier, B.Jakosky for insightful discussions, M.Sorie and the two reviewers for their constructive comments.


%
\appendix
\section{Expression of the sublimation flux of subsurface water ice:}
\label{appendix:subsurfaceflux}
The determination of the sublimation rate of subsurface ice is made using similar arguments to those presented in \citeA{Schorghofer2020}. The flux of vapor in the soil between subsurface ice subliming at depth $z_{\rm{ice}}$~(m) and the surface ($F_{\rm{surface} \longleftrightarrow \rm{surface}}$, in kg~m$^{-2}$~s$^{-1}$) is given by \cite{Hudson2007}:

\begin{equation}
    F_{\rm{subsurface~ice} \longleftrightarrow \rm{surface}} =  \frac{D}{z_{\text{ice}}} \frac{1}{1-\frac{\rho_{\rm{vap,atm}}}{\rho_{\rm{sv}}(T_{\text{ice}})} } \left( \rho_{\rm{sv}}(T_{\text{surf}}) - \rho_{\rm{vap, surf}} \right) 
    \label{appendixeq1}
\end{equation}

\noindent where $\rho_{\rm{vap, surf}}$ is the near-surface vapor density (kg~m$^{-3}$), and the rest of the variables are been defined in section \ref{method:subsurfacesubl}.

The flux of water vapor between the surface and the atmosphere $F_{\rm{surface} \longleftrightarrow \rm{atmosphere}}$, in kg~m$^{-2}$~s$^{-1}$, is given by:

\begin{equation}
    F_{\rm{surface} \longleftrightarrow \rm{atmosphere}} = C_q U \left( \rho_{\rm{vap, atm}} - \rho_{\rm{vap, surf}} \right)
    \label{appendixeq2}
\end{equation}

Mass conservation requires that $F_{\rm{surface} \longleftrightarrow \rm{surface}}$~=~$F_{\rm{surface} \longleftrightarrow \rm{atmosphere}}$. Let's define:
\begin{equation}
    R = \frac{C_q U z_{\rm{ice}}}{D} \left(1-\frac{\rho_{\rm{vap,atm}}}{\rho_{\rm{sv}}(T_{\text{ice}})} \right)
\end{equation}

After some development, one has:

\begin{equation}
    \rho_{\rm{vap, surf}} = \frac{\rho_{\rm{sv}}(T_{\text{ice}}) + R\rho_{\rm{vap,atm}} }{1+R}
\end{equation}

Reinjecting this expression in Eq. \ref{appendixeq1} or \ref{appendixeq2} leads to the expression given in Eq. \ref{eq:fluxgroundice}.

\bibliography{agujournaltemplate.bib}
\end{document}